\documentclass[nohyper,12pt,letterpaper]{JHEP3}
\usepackage[dvips]{epsfig}
\usepackage{amsfonts,amssymb}

\newcommand{\be}{\begin{equation}}
\newcommand{\ee}{\end{equation}}
\newcommand{\bea}{\begin{eqnarray}}
\newcommand{\eea}{\end{eqnarray}}
\newcommand{\ba}{\begin{array}}
\newcommand{\ea}{\end{array}}

\def\bbox{{\,\lower0.9pt\vbox{\hrule \hbox{\vrule height 0.2 cm
\hskip 0.2 cm \vrule height 0.2 cm}\hrule}\,}}
\newcommand{\dsl}{\pa \kern-0.5em /}

\newcommand{\nn}{\nonumber \\}

\newcommand{\EQ}{\begin{equation}}
\newcommand{\EN}{\end{equation}}

\def\bbox{{\,\lower0.9pt\vbox{\hrule \hbox{\vrule height 0.2 cm
\hskip 0.2 cm \vrule height 0.2 cm}\hrule}\,}}

\newcommand{\pa}{\partial}

\font\mybb=msbm10 at 10pt
\def\bb#1{\hbox{\mybb#1}}

\def\bE {\bb{E}}

\title{Quintessence from M-theory}

\author{P.K. Townsend \\
   Department of Applied Mathematics and Theoretical Physics\\
   Centre for Mathematical Sciences \\
   Wilberforce Road, Cambridge CB3 0WA, United Kingdom \\
E-mail: \email{P.K.Townsend@damtp.cam.ac.uk}}

\abstract{The status of exponential scalar potentials in supergravity is
reviewed, and updated. One version of N=8 D=4 supergravity with a positive
exponential potential, obtainable from a `non-compactification' of M-theory,
is shown to have an accelerating cosmological solution that realizes `eternal
quintessence'. Some implications for a de Sitter version of the 
domain wall/QFT correspondence are discussed.}

\keywords{Supergravity, M-Theory, Quintessence}

\begin{document}
\section{Introduction}
\label{intro}

The current evidence for an accelerating early universe can be accomodated
theoretically via a re-introduction of Einstein's (positive) cosmological
constant, which is equivalent to the introduction of tensile matter with
equation of state $P=-\rho$. More generally, and for general spacetime
dimension $D$, tensile matter with equation of state
\be\label{eofs}
P= \kappa \rho
\ee
also yields a flat accelerating universe provided that
\be\label{acc}
-1 \le \kappa < - \left({D-3\over D-1}\right)\, .
\ee
Note that\footnote{We choose units for which $8\pi G=1$.}
\be\label{r00}
R_{00} = {\beta^2\over 2(D-1)} \left[(D-1)P + (D-3)\rho\right]
\ee
where $R_{00}$ is the time-time component of the Ricci tensor, and
\be\label{beta}
\beta = \sqrt{2(D-1)\over (D-2)}\, .
\ee
Acceleration requires negative $R_{00}$, which implies 
a violation of the strong-energy condition. 

One could just set $D=4$, as other more obvious evidence suggests, 
but the observations made in this paper are best understood
in the context of $D$-dimensional supergravity theories for various $D$.
The equation of state (\ref{eofs}) can be realized in a model with a real
scalar field $\phi$ and Lagrangian density
\be\label{lag1}
{\cal L} = \sqrt{-g}\left\{{1\over2} R - (\partial \phi)^2
-V(\phi)\right\}\, ,
\ee
provided that the scalar potential $V$ takes the form
\be\label{exppot}
V= \Lambda e^{-2\alpha\phi}\, ,
\ee
for (positive) dimensionless constant $\alpha$ and positive cosmological
constant 
$\Lambda$ \cite{PR}. As will be shown below, the relation between $\alpha$
and
$\kappa$ in $D$ dimensions (in the conventions used here) is
\be\label{alkap}
\alpha = \beta \sqrt{(1+\kappa)/2}\, .
\ee
Thus $\kappa=-1$ corresponds to $\alpha=0$, a pure cosmological
constant.
As we shall see later, the quantity
\be
\Delta \equiv \alpha^2 -\beta^2
\ee
plays an important role in the supergravity context, and the condition
(\ref{acc}) on $\kappa$ for acceleration translates to
\be\label{deltacon}
-\beta^2 \le \Delta < -2\, .
\ee

A scalar field with a positive potential that yields an accelerating
universe has been called `quintessence' \cite{CDS}; the special 
case under discussion is `eternal quintessence' because 
the fixed equation of state implies an
eternal expansion, exactly as for the original cosmological constant but
with an
adjustable acceleration. From an observational standpoint there is no
particular
merit to eternal quintessence, and other scalar potentials may well be
preferable. However, exponential potentials arise naturally in many
supergravity models, and have some attractive phenomenological features
\cite{BCN}. Moreover, consistent truncations to a single scalar $\phi$ of more
general potentials are usually sums of exponentials, in which case the
potential
$V(\phi)$  will approach a pure exponential for large $|\phi|$. A study of
the
cosmological implications of pure exponential potentials is therefore likely
to
be relevant in quite general circumstances. One aim of this paper is to
examine the implications of supersymmetry for such potentials, in particular
maximal supersymmetry, and to find models that realize either eternal
quintessence or, in the case of potentials that are only asymptotically
exponential, transient quintessence.

As cosmology must ultimately be founded on {\sl quantum} gravity, and as
quantum consistency appears to require string/M-theory, we would wish any
promising cosmological scenario to be derivable from string/M-theory. In
particular, we would wish to be able to embed any D=4 cosmological solution
into some solution of D=11 supergravity or IIB D=10 supergravity. 
However, there is a `no-go' theorem that, subject to certain 
premises, rules out  the possibility of a positive potential, and
hence an accelerating universe, in any effective  D=4 supergravity 
theory obtained in this way \cite{gwg1,nogo}. 
Consider the general case of (warped)
compactification from $D$ to $d<D$ spacetime dimensions on some
compact non-singular manifold of dimension $n=D-d$, and let
$R_{MN}$ and $r_{\mu\nu}$ be the Ricci tensors in $D$ and $d$
dimensions respectively.  The no-go theorem then follows from 
the observation that, for any non-singular D-dimensional 
metric of the form
\be\label{assume}
ds^2_D = f(y) ds^2_d(x) + ds^2_n(y)\, ,
\ee
positivity of $R_{00}$ implies positivity of $r_{00}$. In other words,
for compactifications of the assumed form the strong energy 
condition in spacetime dimension $D$ implies 
the strong energy condition in spacetime dimension $d<D$. 
The latter condition is equivalent to $|g_{00}|V \le (d-2)\dot\phi^2$, and
hence $V\le0$ if initial conditions can be chosen such that
$\dot\phi=0$. More directly, as follows from (\ref{r00}) with
$D\rightarrow d$, the $d$-dimensional strong energy condition
forbids an accelerating $d$-dimensional universe.
As recently emphasized \cite{HKS,FKMP}, 
this makes it difficult to embed accelerating cosmologies 
into string/M-theory because the strong energy condition is
satisfied by both D=11 supergravity and IIB D=10 supergravity.

 The strong energy condition in D-dimensions may be violated by 
a $(D-2)$-form gauge potential \cite{gwg1}. D=11 supergravity
compactified on any Ricci flat $7$-manifold has such a field, and
a modified $T^7$-compactification that exploits this was shown in 
\cite{ANT} to yield a `massive' D=4 N=8 supergravity with a 
positive exponential potential \cite{ANT}. However, this
model has $\Delta=4$ and so will not yield an accelerating universe. 
The strong energy condition may also be
violated by scalar field potentials. This possibility is realized in
some D=4 N=4 supergravity theories with de Sitter (dS) vacua \cite{GZ};
these models are truncations of certain non-compact gaugings of N=8 
supergravity \cite{Hull2} which are obtainable by `compactification'
of D=11 supergravity \cite{HW}. The no-go theorem is evaded in this
case \cite{Hull1} because the `compactifying' space is actually 
non-compact (see \cite{CL} for another type of `non-compactification' 
to de Sitter space). As shown in \cite{HW}, similar `non-compactifications'
yield all the non-compact gauged maximal supergravity theories in
D=4,5,7  \cite{GRW,PPvN}. 
 
Here it will be shown that a particular non-compact gauged
N=8 D=4 supergravity has a positive exponential potential satisfying the
condition (\ref{deltacon}) needed for an accelerating cosmology. 
This model provides a realization of eternal quintessence 
with equation of state $P = -(7/9)\, \rho$; as will be explained later,
it is obtainable from a warped `non-compactification' of
M-theory, but it  can also be obtained from a similar
`non-compactification' of IIB superstring theory on $H^{(3,3)}\times
S^1$, where $H^{(p,q)}$ are the  $(p+q-1)$-dimensional non-compact
spaces described in \cite{HW}. The $H^{(3,3)}$ `compactification' yields 
the $SO(3,3)$ gauged D=5 maximal supergravity of \cite{GRW}, and the
D=4 model is obtained by a further dimensional reduction on $S^1$. Moreover,
{\it the accelerating D=4 cosmology is the $S^1$ reduction of D=5 dS
space}; this has some implications for the non-conformal generalization of
the dS/CFT correspondence \cite{hull3,strom} that will be mentioned in the
concluding comments. 

Of course, as long as the physical acceptability of `non-compactifications'
remains dubious, neither this example of eternal quintessence nor the dS
cases
discussed in \cite{Hull1} can settle the question of whether string/M-theory
is
compatible with an accelerating universe. This and other points
arising from the results found here will be discussed in a final section.
For the moment we turn to a study of the implications of quintessence 
for D-dimensional isotropic and homogeneous cosmologies.

\section{D-dimensional Quintessence}

Consider a D-dimensional FRW spacetime with scale factor $a(t)$ and
metric
\be
ds^2 = -dt^2 + a(t)^2 d\Omega^2\, ,
\ee
where $d\Omega^2$ is the metric of the maximally-symmetric $(D-1)$-space with
curvature $k^{-1}$, for $k=-1,0,1$. 
If this spacetime is filled with a perfect fluid of mass density $\rho$ and
pressure $P$ then the continuity equation for the fluid is
\be\label{cont}
\dot\rho = -(D-1)(\rho + P)\left({\dot a\over a}\right)\, ,
\ee
while the Friedmann equation for the scale factor is
\be\label{Feq}
\left({\dot a\over a}\right)^2 -{2\rho\over (D-1)(D-2)} = - {k\over a^2}\, .
\ee
These two equations imply that
\be\label{acc2}
(D-1)(D-2) \ddot a = - a\left[(D-3)\rho + (D-1)P\right]\, ,
\ee
and hence an accelerating universe when
\be\label{ac1}
P < - {(D-3)\over (D-1)}\, \rho\, .
\ee
For the equation of state (\ref{eofs}) this translates to the condition
(\ref{acc}) on $\kappa$. Also, using (\ref{eofs}) in (\ref{cont})
we deduce that
\be\label{rrho}
\rho \propto a^{-(D-1)(1+\kappa)}\, .
\ee

Now consider a single scalar model with arbitrary potential $V$. The scalar
equation is (see e.g. \cite{GT})
\be\label{scalareq}
\ddot \phi + (D-1)\left({\dot a\over a}\right)\dot\phi + \frac12{\partial
V\over \partial\phi}=0\, ,
\ee
while the Friedmann equation is (\ref{Feq}) with
\be
\rho = \dot\phi^2 + V\, .
\ee
Consistency of these equations with the continuity equation (\ref{cont})
implies that
\be
P= \dot\phi^2 -V\, ,
\ee
and hence that
\be
\dot\phi^2 = {1\over2}(\rho +P)\, ,\qquad V={1\over2}(\rho -P)\, .
\ee
Given the equation of state (\ref{eofs}) we then have, in particular, that
\be
V= {1\over2}(1-\kappa) \rho\, ,
\ee
and hence from (\ref{rrho}) that
\be\label{veer}
V\propto (1-\kappa) a^{-(D-1)(1+\kappa)}\, ,
\ee
with a positive constant of proportionality. Similarly, we also have that
$\dot\phi^2 = (1+\kappa)\rho/2$ or, equivalently,
\be
\rho = {2\dot\phi^2\over 1+\kappa}\, .
\ee
Using this in (\ref{Feq}), and setting $k=0$, we deduce that
\be
a^{-(D-1)(1+\kappa)} \propto e^{-\beta\sqrt{2(1+\kappa)}\, \phi}\, ,
\ee
and hence, from (\ref{veer}), that\footnote{Here we use $(1-\kappa)=
-2(\alpha^2-\beta^2)/\beta^2$. The $\alpha=\beta$ case is 
special and must be dealt with separately; one finds that no power-law
solutions are possible for any $k$.}
\be\label{vee}\label{Vdelta}
V\propto -\Delta\,  e^{-2\alpha\phi}\, ,
\ee
with a positive constant of proportionality, and with 
$\alpha$ related to $\kappa$ by (\ref{alkap}).

As a check, we look for power-law solutions of the scalar and
Friedmann equations of the form
\be
a= t^\gamma \, ,\qquad e^{\alpha\phi} =t
\ee
for constant $\gamma$. The scalar equation (\ref{scalareq}) is solved
for $V$ of the form (\ref{exppot}) if
\be
(D-1)\gamma = (1+ \alpha^2\Lambda/2)\, .
\ee
Using this, the Friedmann equation is solved for $k=0$ if
\be
\Lambda = - 2\alpha^{-4}\Delta\, ,
\ee
in agreement with (\ref{vee}). It then follows that
\be
\gamma = {2\over \alpha^2 (D-2)}\, , 
\ee
and the condition $\gamma>1$ for acceleration is thus seen to be equivalent
to $\Delta <-2$, as expected.  For future purposes we remark that
$\gamma=3$ for $D=4$ and $\alpha=1/\sqrt{3}$.

\section{Quintessence in N=1 supergravity}

Scalar fields of minimal D=4 supergravity models necessarily belong to 
chiral
supermultiplets. As we are interested in a single scalar field we need
consider
only a single chiral multiplet, for which the physical bosonic fields
consist of
one scalar $\phi$ and one pseudoscalar, although to get the general
potential
for $\phi$ we will also need to include one vector multiplet. Given that
the coupling of supergravity to one chiral supermultiplet and 
one vector supermultiplet preserves
parity\footnote{If the superpotential is such as to violate parity
then the truncation to the single scalar $\phi$ may not be consistent,
in which case the arguments to follow may require modification. See
\cite{BBS} for a discussion of `two-field' quintessence models.}, 
we may consistently set to zero the pseuodoscalar in the chiral
multiplet.
Having done so, we arrive at the bosonic Lagrangian
\be\label{lag2}
{\cal L} = \sqrt{-g}\left\{{1\over2} R - {1\over2} h^{-1}(\phi) F^2 -
(\partial
\phi)^2  -V(\phi)\right\}\, ,
\ee
where the $F^2$ term is the kinetic term for the vector field of the vector
multiplet, which is coupled to the scalar field through the real function
$h(\phi)$. The scalar potential $V$ takes the form
\be\label{genpot}
V = (w')^2 - \beta^2 w^2 + \xi^2 h
\ee
where $w(\phi)$ is a real superpotential\footnote{Initially $w$ is a complex
function of the chiral superfield, but after the truncation of the
pseodoscalar in the chiral multiplet no generality is lost by restricting it
to be a real function.} and $\xi$ is a Fayet-Illiopoulos (FI) constant. The
superpotential terms are those derived in \cite{pkt1} for general spacetime
dimension $D$ on the assumption of positive energy and the existence of an
adS vacuum of a certain type. The latter condition can be relaxed, but the
existence
of the last term, the `D-term' potential, of (\ref{genpot}) shows that more
general potentials for $\phi$ are possible, at least in D=4; this is not
immediately obvious because a given function $h$ might be
equivalent to a modification of the superpotential $w$, but the examples to be
considered below show that this is not so in general.

If we seek a pure exponential potential then we must choose
\be
w= me^{-\alpha \phi}\, ,\qquad h= e^{-2\alpha\phi}\, ,
\ee
for mass parameter $m$. In this case
\be\label{pot2}
V= \Lambda e^{-2\alpha\phi}
\ee
with
\be
\Lambda = m^2(\alpha^2 -\beta^2) + \xi^2 \, .
\ee
When $\xi=0$ we have $\Lambda= \Delta$, so in this case the conditions
$\Delta
<-2$ and $\Lambda>0$ that are needed for an accelerating universe are
incompatible \cite{HKS}. But if $\xi\ne0$ we can easily arrange for both
these conditions to be satisfied \cite{BD,Kallosh}. 
As we shall see later, this possibility is realized by
both `massive' and gauged versions of N=8 D=4 supergravity.

In the absence of the FI term the choice $\alpha=\beta$ is rather special
because it leads to a vanishing potential despite a non-vanishing
superpotential. This fact allows the following possibility. Consider the
superpotential
\be
w = m\left[e^{-\beta\phi} -e^{-\alpha\phi}\right]
\qquad (0<\alpha <\beta).
\ee
This yields the potential
\be
V= ( \beta-\alpha) m^2\left[2\beta e^{-(\alpha+\beta)\phi} -
(\alpha+\beta) e^{-2\alpha\phi}\right]\, .
\ee
As $\alpha<\beta$, the first term dominates as $\phi\rightarrow
-\infty$,
and we then have an accelerating universe if
\be
\alpha < 2\sqrt{\beta^2-2} -\beta\, .
\ee
In the $\phi\rightarrow \infty$ limit the second term in the potential
vanishes
and we have
\be
V \sim -(\beta^2-\alpha^2) e^{-2\alpha\phi}\, .
\ee
As $V$ is now negative the universe is decelerating. We thus have a period
of
acceleration for large negative $\phi$, with a transition to deceleration as
$\phi$ passes through zero; the universe then rolls towards $V=0$ from
below.

\section{Constraints from extended supersymmetry}

As pointed out in \cite{LPSS}, the variable $\Delta=\alpha^2-\beta^2$ has
the
property that it is unchanged by dimensional reduction, if after the
reduction
one then consistently truncates to a model with a single scalar field.
For this reason it was used in \cite{LPT} as the basis of a classification of
exponential potentials in maximal and half-maximal supergravity
theories; the values of $\Delta$ identified as occuring 
in some massive or gauged supergravity fall into the following four classes:
\be
\Delta  <-2,\qquad  -2 \le \Delta <0 ,\qquad  \Delta =0, \qquad
\Delta >0\, .
\ee
We shall comment on each of these cases in turn, updating the discussion of
\cite{LPT} where appropriate:

\begin{itemize}

\item $\Delta <-2$. This case, which was the focus of \cite{LPT}, is
realized by toroidal reductions of the gauged maximal supergravity theories
in
D=3,4,5 and 7, and also of the gauged `F(4) supergravity' in D=6. This is
because these theories admit adS vacua and hence a truncation to a theory
without scalars, or equivalently to a theory with a scalar having a constant
negative potential, corresponding to $\alpha=0$ and $\Lambda <0$.
The values of $\Delta$ found this way are all such that $\Delta<-2$.

Of relevance here is the fact, not discussed in \cite{LPT}, that a very
similar
analysis can be made for those (D=4,5) {\it non-compact} gauged supergravity
theories
that admit dS vacua. For exactly the same reasons, these theories can be
consistently truncated to a theory without scalars, and a subsequent
toroidal
reduction and consistent truncation again yields a single scalar model with
$\Delta<-2$ but now with a {\it positive} cosmological 
constant $\Lambda$. We will
consider one such case in more detail in the following section.

\item $-2 \le \Delta <0$. This case is realized by several gauged
supergravity
theories without adS vacua, and probably by other truncations of those with
adS
vacua. All known examples have $\Delta=-2$, except the maximal gauged D=8
supergravity theory (and hence its toroidal reductions) for which the 
obvious single scalar truncation yields $\Delta =-4/3$.

\item $\Delta=0$. This was the `fourth type' in the classification
of \cite{LPT}; as stated there, there are no known supergravity
examples\footnote{The published version of this paper erroneously puts
the massive N=8 supergravity of \cite{ANT} into this category; 
for reasons explained below, it belongs to a separate category.}, which may be related to the fact that
the formula (\ref{Vdelta}) gives zero potential for $\Delta=0$.

\item $\Delta >0$. These are the `massive' supergravity theories with
positive
potential. The prototype is the massive IIA D=10 theory, for which
$\Delta=4$. 
Many massive $D\le9$ theories with $\Delta=4/N_c$, for integer $N_c$, were
obtained in \cite{CLPST} by non-trivial dimensional reduction from D=11
($N_c=1,2,3,4$ for D=4). All these theories have supersymmetric domain
wall `vacua' in which the only singularity is a delta-function singularity
in
the curvature at the location of the walls.

\end{itemize}

The massive N=8 supergravity of \cite{ANT} mentioned earlier could be put into 
the $\Delta>0$ category but, for present purposes at least, it is better to consider it as
belonging to a separate category. The general construction 
starts from a $D$-dimensional theory with  Lagrangian density
\be
{\cal L}_D = \sqrt{-g}\left\{ {1\over2}R - {1\over (p+2)!}
F^2_{(p+2)}\right\}\, , 
\ee
where $F_{(p+2)}=dA_{(p+1)}$ is a (p+2)-form field strength for the
(p+1)-form gauge potential $A_{(p+1)}$.  Let us consider a compactification to (p+2)
dimensions, with a reduction/truncation ansatz for which $F_{(p+2)}$ is
restricted to the $(p+2)$-dimensional spacetime and the
$D$-metric takes the form
\be
ds^2_D =  e^{-2a\phi} ds^2_{p+2}+
e^{2b\phi}ds^2(T^{D-p-2})\nn\, , 
\ee
where $(D-p-2)b= pa$. For the choice
\begin{equation}\label{achoice}
a= \sqrt{\frac{2(D-p-2)}{(D-2)p}}\, , 
\end{equation}
the Lagrangian density governing the
dynamics of the $(p+2)$-dimensional fields is
\be
{\cal L}_d = \sqrt{-g}\left\{ {1\over2}R - (\partial \phi)^2 -
{1\over (p+2)!} e^{2(d-1)a \phi} F^2_{(p+2)}\right\}\, .
\ee
The $A_{(p+1)}$ field equation is almost trivial but its general solution
introduces a mass parameter $m$ as an integration constant. Taking this into
account, an equivalent Lagrangian density for the other fields is
\be\label{Ld}
{\cal L}_d = \sqrt{-g}\left\{ {1\over2}R - (\partial \phi)^2 - m^2
e^{-2\alpha\phi}\right\}\, , 
\ee
where $\alpha = (d-1)a$. Using (\ref{achoice}) and (\ref{beta}), one finds that 
\begin{equation}
\Delta = \alpha^2 -\beta^2 = \frac{2(p+1)(D-p-3)}{D-2} >0\, . 
\end{equation}
In particular $\Delta=4$ for $D=11$ with $p=2,5$ (and for $D=10$ with $p=3$). However,  the formula (\ref{Vdelta}) 
implies a negative potential for $\Delta>0$, whereas the potential of (\ref{Ld}) is positive,
for reasons explained in  \cite{ANT}. 

\section{Eternal Quintessence from de Sitter spacetimes}

Of all the cases enumerated above only the non-compact gauged supergravity
theories with dS vacua, and their dimensional reductions, satisfy the two
conditions, $\Lambda>0$ and $\Delta <-2$, required for an accelerating
universe. There may be other possibilities in supergravity theories with
fewer
supersymmetries; we have seen that the condition for acceleration is not
difficult to satisfy within N=1 D=4 supergravity, even for a pure
exponential
potential, and more general potentials are of course possible. However, it
is unclear how these other cases would arise from M-theory, whereas it is
known
how all maximal supergravity theories are related to M-theory. The
non-compact
N=8 D=4 supergravity theories with dS vacua have been recently reviewed in
the 
context of the issues discussed here; these provide accelerating cosmologies
with equation of state $P=-\rho$. Here we wish to show that another of the
non-compact gaugings of N=8 supergravity provides a realization of eternal
quintessence. 

One way to obtain this model is to start from the D=5 $SO(3,3)$
gauged supergravity, which has a D=5 adS vacuum \cite{GRW}. 
Reduction to D=4 yields the
$CSO(3,3,2)$ gauged N=8 supergravity (in the notation of \cite{Hull2}). As
explained earlier, this model has a consistent truncation to a single scalar
field with $\Delta = -8/3$ because this corresponds to $\Delta=-\beta^2$ in
D=5, and $\Delta$ is unchanged by dimensional reduction. This yields the
equation of state
\be\label{eofs2}
P= -(7/9)\, \rho
\ee
and hence acceleration. 

Some aspects of this model can be understood as follows. We
saw that an accelerating universe can only be obtained from an exponential
potential
in N=1 supergravity by inclusion of a FI term. In this case, the coefficient
$\alpha$ can be identified as the dilaton coupling constant. Now, the only
values of the dilaton coupling constant that occur in consistent truncations
of
N=8 supergravity to a theory with a single scalar are such that
$\alpha^2 = 0,{1\over3}, 1, 3$  \cite{HT}. 
Moreover, if we start from a theory in D=5 without scalars and reduce to D=4
then only the values $1/3$ and $3$ of $\alpha^2$ occur; the first case corresponds precisely
to $\Delta= -8/3$ and hence to $\kappa =-7/9$ (the other case corresponds to
$\Delta=0$, but then the potential is zero). 
The $\alpha^2= 1$ case corresponds to $\Delta=-2$ but 
this does not yield an accelerating universe. Note that if a 
dS vacuum were possible for $D>5$ it would yield other values of $\Delta$ corresponding to
disallowed values of $\alpha$ in $D=4$; this provides another way to see why
the non-compact gaugings of $D\ge6$ supergravity theories 
do not admit dS vacua.

We will now verify that the accelerating D=4 universe with equation of state
(\ref{eofs2}) lifts to D=5 de Sitter space. We start with the $D=5$
Lagrangian density
\be\label{d5}
{\cal L}_5 = \sqrt{-\det g^{(5)}} \left[ R^{(5)} + \Lambda \right]\, .
\ee
If the 5-metric is written as
\be
ds^2_5 = e^{-(2/\sqrt{3})\phi(x)} ds^2_4(x) + 
e^{(4/\sqrt{3})\phi(x)}dy^2\, ,
\ee
where $x$ denotes the 4-space coordinates and $y$ parameterizes a circle,
then
the 4-metric and scalar $\phi$ are governed by a $D=4$ Lagrangian density of
the
form (\ref{lag1}) with
\be
V= \Lambda e^{-(2/\sqrt{3})\phi}
\ee
and hence $\alpha=1/\sqrt 3$, as claimed earlier. The accelerating
4-dimensional universe was discussed in section 2 (where it was shown
to have $a=t^3$); the 4-metric and dilaton are
\be\label{4metdil}
ds^2_4 = -dt^2 + t^6 ds^2(\bE^3)\, ,\qquad \phi = \sqrt{3}\log t\, .
\ee
The corresponding 5-metric is
\be
ds^2_5 =  -(d \log t)^2 + t^4 ds^2(\bE^4)\, ,
\ee
which is $dS_5$ space in planar-type coordinates; as such it is obviously a
solution of the D=5 theory with Lagrangian density (\ref{d5}). It will
also be a solution of the $SO(3,3)$ gauged maximal D=5 supergravity,
but presumably one at a saddle point of the potential rather than a
maximum. This will mean that the dS solution is unstable, but this
instability may be a good thing in that it allows an escape from 
eternal acceleration. 

What remains to be explained about the D=4 supergravity model with
this accelerating cosmological solution is its relation to M-theory.
As mentioned in the introduction, the D=5 $SO(3,3)$ gauged
supergravity can be obtained
from a warped `compactification' of IIB supergravity on
$H^{(3,3)}$. However, after the further reduction to $D=4$ on $S^1$, we
may pass to the dual IIA supergravity (or, more accurately,
superstring theory). As this is an $S^1$ compactification of M-theory
we now have M-theory `compactified' on $H^{(3,3)}\times T^2$, which
yields \cite{HW} the non-compact gauging of N=8 D=4 supergravity with gauge
group $CSO(3,3,2)$; this is the same
theory as one obtains by reduction of the $SO(3,3)$ gauged D=5
supergravity.

\section{Comments}

One premise of the no-go theorem of \cite{gwg1,nogo} is the
time-independence of the compactifying space. This condition 
is violated by the accelerating D=4 cosmological solution just
discussed, so it was {\it a priori} possible for
this solution to arise from a true compactification of D=11
supergravity. The fact that it does not (because of the non-compactness
of the internal space) is evidently related to its 5-dimensional interpretation
as periodically-identified de Sitter space.

One point that should be appreciated is that the conclusion of the
no-go theorem that $V\le0$ does not mean that the function $V(\phi)$
must be everywhere non-positive. It is typically the case that $V$ is both
unbounded from below {\it and from above} in compactifications of
D=11 supergravity that satisfy all the premises of the
theorem. Rather, the theorem states that the {\it value} of $V$ in any
solution of the assumed form is non-positive. Generically, there will
be directions in field space for which $V$ is positive and
exponentially increasing as one goes to infinity. There are
power-law cosmological solutions associated with this limit, 
with positive $V$ (provided that $\Delta \ne0$). This does not 
contradict the no-go theorem because it says nothing about this type of 
cosmological compactification. Nevertheless, the only 
{\it accelerating} D=4 cosmological solution that the author has been
able to find in this way is the one described above that is related to
M-theory by a `non-compactification'. 

It should also be borne in mind that string/M-theory actually goes
beyond supergravity in that it involves branes sources. The equation
of state for a gas of p-branes is \cite{ABE}
\be
\kappa = -{p\over (D-1)}\, ,
\ee
which leads to acceleration for $p\ge D-2$. The $p=D-1$ case is a
space-filling brane and its tension adds to the cosmological
constant, but if the extra dimensions are compact then the sum
over the tensions of all branes must vanish. The $p=D-1$ case
corresponds to a gas of domain walls for $D=4$ \cite{BBS2},
although it is not clear whether this solves the problem.  

Finally, it is instructive to compare the $S^1$ reduction of dS space to
the $S^1$ reduction of {\it anti} de Sitter (adS) space considered in
\cite{LPT}; the latter results in a domain wall (DW) solution of the lower
dimensional theory. The domain wall metric in the `dual frame' is again
adS but the dilaton resulting from the compactification breaks the adS
group to the Poincar\'e group on the DW; the 
adS/CFT correspondence is therefore
replaced with a DW/QFT correspondence \cite{BST}. To apply these ideas
to the reduction of $dS_5$ discussed above, we again introduce a new 
`dual-frame' 4-metric
\be
d\tilde s_4^2 \equiv e^{-(2/\sqrt{3})\phi} ds^2_4 \nn
= -d(\log t)^2 + t^4 ds^2(\bE^3)\, .
\ee
This is $dS_4$ but the SO(4,1) invariance is broken to $ISO(3)$ 
by the dilaton. Thus, the dS/CFT correspondence
\cite{hull3,strom} is replaced in the non-conformal case by a 
correspondence between a cosmology and a Euclidean field theory, with
the p-brane of the DW/QFT correspondence being replaced by a Cauchy
surface of the cosmological spacetime. A change in scale shifts the
dilaton and hence moves the Cauchy surface in time.

\acknowledgments
I thank Robert Brandenberger, Martin Bucher, Gary Gibbons, 
Chris Hull, Nemanja Kaloper, Carlos Nu\~nez and
Fernando Quevedo for helpful conversations.

\newcommand{\NP}[1]{Nucl.\ Phys.\ {\bf #1}}
\newcommand{\AP}[1]{Ann.\ Phys.\ {\bf #1}}
\newcommand{\PL}[1]{Phys.\ Lett.\ {\bf #1}}
\newcommand{\CQG}[1]{Class. Quant. Gravity {\bf #1}}
\newcommand{\CMP}[1]{Comm.\ Math.\ Phys.\ {\bf #1}}
\newcommand{\PR}[1]{Phys.\ Rev.\ {\bf #1}}
\newcommand{\PRL}[1]{Phys.\ Rev.\ Lett.\ {\bf #1}}
\newcommand{\PRE}[1]{Phys.\ Rep.\ {\bf #1}}
\newcommand{\PTP}[1]{Prog.\ Theor.\ Phys.\ {\bf #1}}
\newcommand{\PTPS}[1]{Prog.\ Theor.\ Phys.\ Suppl.\ {\bf #1}}
\newcommand{\MPL}[1]{Mod.\ Phys.\ Lett.\ {\bf #1}}
\newcommand{\IJMP}[1]{Int.\ Jour.\ Mod.\ Phys.\ {\bf #1}}
\newcommand{\JHEP}[1]{J.\ High\ Energy\ Phys.\ {\bf #1}}
\newcommand{\JP}[1]{Jour.\ Phys.\ {\bf #1}}


\begin{thebibliography}{99}

\bibitem{PR}
B. Ratra and P.J.E. Peebles, {\sl Cosmological consequences of a rolling
homogeneous scalar field}, Phys. Rev. {\bf D37} (1998) 3406.

\bibitem{CDS}
R.R. Caldwell, R. Dave and P.J. Steinhardt, {\sl Cosmological imprint
of an energy component with general equation of state},
Phys. Rev. Lett. {\bf 80} (1998) 1582.

\bibitem{BCN}
T. Barreiro, E.J. Copeland, N.J. Nunes, {\sl Quintessence arising from
exponential potentials}, Phys. Rev. {\bf D61} (2000) 127301. 

\bibitem{gwg1}
G.W. Gibbons, {\sl Aspects of Supergravity Theories}, in
{\it Supersymmetry, Supergravity, and Related Topics}, eds.
F. del Aguila, J.A. de Azc{\' a}rraga and L.E. Iba{\~ n}ez (World
Scientific 1985) pp. 346-351.

\bibitem{nogo}
J. Maldacena and C. Nu\~nez, {\sl Supergravity description of field
theories on curved manifolds and a no-go theorem},
Int. J. Mod. Phys. {\bf A16} (2001) 822.

\bibitem{HKS}
S. Hellerman, N. Kaloper and L. Susskind, {\sl String Theory and
Quintessence}, JHEP {\bf 0106} (2001) 003. 

\bibitem{FKMP}
W. Fischler, A. Kashani-Poor, R. McNees and S. Paban, {\sl The
acceleration of the universe, a challenge for String Theory},
JHEP {\bf 0107} (2001) 003.  

\bibitem{ANT}
A. Aurilia, H. Nicolai and P.K. Townsend, {\sl Hidden constants:the
$\theta$-parameter of QCD, and the cosmological constant of N=8
supergravity}, 
Nucl. Phys. {\bf B176} (1980) 509.

\bibitem{GZ}
S.J.Gates, Jr. and B. Zwiebach, {\sl Gauged N = 4 Supergravity Theory
with a New Scalar Potential}, Phys. Lett. {\bf 123B} (1983) 200;
{\sl Searching for all N = 4 Supergravities in Superspace}, 
Nucl. Phys. {\bf B238} (1984) 99. 

\bibitem{Hull2}
C.M. Hull, {\sl A new gauging of N=8 supergravity}, Phys. Rev. {\bf D30}
(1984) 760; {\sl Non-compact gaugings of N=8 supergravity}, Phys. Lett. {\bf
142B} (1984) 39; {\sl More gaugings of N=8 supergravity}, Phys. Lett. {\bf
148B} (1984) 297; {\sl The minimal couplings and scalar 
potentials of the gauged N=8
supergravities}, Class. Quantum Grav. {\bf 2} (1985) 343.

\bibitem{HW}
C.M. Hull and N.P. Warner, {\sl Non-compact gaugings from higher
dimensions}, Class. Quantum Grav. {\bf 5} (1988) 1517.

\bibitem{Hull1}
C.M. Hull, {\sl De Sitter space in supergravity and M-theory},
hep-th/0109213. 

\bibitem{CL}
A. Chamblin and N.D. Lambert, {\sl de Sitter space from M-theory}, 
Phys. Lett. {\bf 508B} (2001) 369. 

\bibitem{GRW}
M. G\"unaydin, L.J. Romans and N.P. Warner, {\sl Compact and non-compact
gauged
supergravity theories in five dimensions}, Nucl. Phys. {\bf B272} (1986)
598.

\bibitem{PPvN}
M. Pernici, K. Pilch and P. van Nieuwenhuizen, {\sl Gauged N=8 D=5
supergravity}, Nucl. Phys. {\bf B259} (1985) 460;
{\sl Noncompact gaugings and
critical points of maximal supergravity in seven dimensions},
Nucl. Phys. {\bf B249} (1985) 381.

\bibitem{hull3}
C.M. Hull, {\sl Timelike T-duality, de Sitter space, large N gauge
theories and topological field theory}, JHEP {\bf 9807} (198) 021. 

\bibitem{strom}
  A.~Strominger,
  ``The dS / CFT correspondence,''
  JHEP {\bf 0110} (2001) 034.

 \bibitem{GT}
G.W. Gibbons and P.K. Townsend, {\sl Cosmological evolution of degenerate
vacua}, Nucl. Phys. {\bf B282} (1987) 610.

\bibitem{BBS}
  M.~C.~Bento, O.~Bertolami and N.~M.~C.~Santos,
  ``A Two field quintessence model,''
  Phys.\ Rev.\ D {\bf 65} (2002) 067301.

\bibitem{pkt1}
P.K. Townsend, {\sl Positive energy and the scalar potential in higher
dimensional (super)gravity theories}, Phys. Lett. {\bf 148B} (1984)
55.

\bibitem{BD}
P. Bin\'etruy and G. Dvali, {\sl D-term inflation}, Phys. Lett.
{\bf 388B} (1996) 241.

\bibitem{Kallosh}
R. Kallosh, {\sl N=2 supersymmetry and de Sitter space},
hep-th/0109213.

\bibitem{LPSS}
H. Lu, C.N. Pope, E. Sezgin and  K.S. Stelle, {\sl Stainless super
p-branes}, 
Nucl. Phys. {\bf B456} (1995) 669; {\sl Dilatonic p-brane solitons},
Phys. Lett. {\bf 371B} (1996) 46.

\bibitem{LPT}
H. L{\" u}, C.N. Pope and P.K. Townsend, {\sl Domain walls from anti-de
Sitter
space}, Phys. Lett. {\bf 391B} (1997) 39.

\bibitem {CLPST}
P.M. Cowdall, H. L{\" u}, C.N. Pope, K.S. Stelle and P.K. Townsend, {\sl
Domain
walls in massive supergravities}, Nucl. Phys. {\bf B486} (1997) 49.

\bibitem{HT}
  C.~M.~Hull and P.~K.~Townsend,
  Nucl.\ Phys.\ B {\bf 438} (1995) 109.
  
\bibitem{ABE}
S. Alexander, R. H. Brandenberger and D. Easson, {\sl Brane gases in 
the early universe}, Phys. Rev. {\bf D62} (2000) 103509. 
 
\bibitem{BBS2}
R.A. Battye, M. Bucher and D. Spergel, {\sl Domain wall dominated
universes}, astro-ph/9908047. 

\bibitem{BST}
H.J. Boonstra, K. Skenderis and P.K. Townsend, {\sl The domain-wall/QFT
correspondence}, JHEP {\bf 9901} (1999) 003.




    

\end{thebibliography}
\end{document}